# Novel Runtime Systems Support for Adaptive Compositional Modeling on the Grid

Srinidhi Varadarajan and Naren Ramakrishnan
Department of Computer Science
Virginia Tech, Blacksburg, VA 24061, USA
Email: srinidhi@cs.vt.edu, naren@cs.vt.edu

**Abstract**
Grid infrastructures and computing environments have progressed significantly in the past few years. The vision of truly seamless Grid usage relies on runtime systems support that is cognizant of the operational issues underlying grid computations and, at the same time, is flexible enough to accommodate diverse application scenarios. This paper addresses the twin aspects of Grid infrastructure *and* application support through a novel combination of two computational technologies – *Weaves*, a *source-language independent* parallel runtime compositional framework that operates through reverse-analysis of compiled object files, and runtime recommender systems that aid in dynamic knowledge-based application composition. Domain-specific adaptivity is exploited through a novel compositional system that supports runtime recommendation of code modules and a sophisticated checkpointing and runtime migration solution that can be transparently deployed over Grid infrastructures. A core set of "adaptivity schemas" are provided as templates for adaptive composition of large-scale scientific computations. Implementation issues, motivating application contexts, and preliminary results are described.

## 1 Introduction

Grid computing [Berman et al., 2003] is increasingly becoming a reality and rapid advances are being made to establish high performance software environments for scientific and engineering computations. In particular, there has been a recent shift of emphasis from low-level application scheduling and execution to creating infrastructure for high-level problem solving environments (PSEs) or grid computing environments (GCEs) [Fox et al. 2003]. To be effective, such GCEs should provide high-level, powerful, computational primitives [Lee and Talia, 2003] within the context of the emerging landscape of Grid infrastructures. This requires both an understanding of the architectural assumptions of computational grids and an appreciation for how disciplinary scientists do computational science.

In consideration of the target of this special issue, this paper focuses on the twin aspects of Grid infrastructure *and* application support, especially toward a unifying framework that addresses issues pertinent to both aspects. The focus is on **runtime systems support** that is cognizant of the operational issues underlying grid computations and is flexible enough to accommodate diverse application scenarios. We begin by identifying specific desiderata for runtime systems support on the Grid.

*Infrastructural and Usage Considerations:* The vision of Grid usage is to tap into a vast computational resource, with a reliability reminiscent of yesteryears custom designed supercomputers. Today's Grid infrastructure, however, resembles a loosely organized cluster of clusters, with all its attendant shortcomings. Some of these shortcomings are borne out by the experiences of one of the authors, who has set up and managed a 200 node cluster, operational for over two years, and supporting a large number of demanding computational science applications. For instance, the probability of failure of an arbitrary cluster node involved in a large computation is relatively high, and grows dramatically with increasing cluster sizes.[1] In addition, we believe that while the initial Grid infrastructure will consist of a small number of large supercomputing facilities, the Grid will evolve to include large numbers of relatively small clusters. In such a scenario, there will be a constant tussle between the ability of large jobs to run effectively and small players being able to assert administrative control over their own resources. We argue that fundamental Grid reliability and resource control issues need to be resolved through runtime systems support to make Grid computing an everyday reality.

*Software Engineering Aspects:* Scientific applications emerging on nascent Grid infrastructures are expected to handle issues stemming from the enormous heterogeneity of Grid platforms – heterogeneity of architecture, of processor speeds, and of interconnection bandwidth. While newer generations of scientific software can be designed with this consideration in mind, targeting large legacy scientific codes for the Grid poses an almost impossible software engineering endeavor. What is needed is runtime systems support that can effectively *mask* the complexity of Grid infrastructures, to enable this transition.

*Adaptivity and Modeling for Grid Applications:* When basic infrastructural and software engineering issues are resolved, Grid computing holds promise for scientific codes to become more adaptive - adaptive in terms of algorithm selection, architectural tuning, and exploiting the underlying scientific usage contexts [Foster et al., 2001]. We posit a broad picture of adaptivity here, one which is not restricted to identifying partitioning parameters, modifying data decompositions, or parallel scheduling; instead, adaptivity is proposed at a more logical unit of algorithms and object codes. This viewpoint leads to scientific codes being organized in a model-based framework for adaptive composition, execution, and performance analysis. As Berman et al. point out [Berman et al., 2003], "we still cannot throw any application at the Grid and have resource management software determine where and how it will run."

These considerations lead us to identifying important requirements for runtime systems support. First, runtime systems support should enable a transparent transition path for composing and executing legacy codes, without requiring that they be rewritten to achieve this functionality. Second, runtime adaptivity should allow the dynamic selection, reconfiguration, and execution of code modules, taking into account performance considerations and problem characteristics. Third, automatic facilities for masking systems failures should be provided (witness the recent thrust for "recovery-oriented computing" initiated by Patterson [Patterson et al., 2002]). Finally, runtime systems support is needed to realize a Grid infrastructure that reconciles the needs of large computational applications and administrative control, through a fluidic definition and control of Grid resources.

## *Solution Approach*

Our solution approach for runtime systems support is two-pronged: (i) domain-specific adaptivity is exploited through a novel compositional system that supports runtime recommendation of code modules; and (ii) a sophisticated checkpointing and runtime migration solution is provided for deployment over the Grid. A core set of "adaptivity schemas" constitute a reconfigurable approach to steering and managing large-scale scientific computations.

In this view of grid computing, a high-level problem specification (e.g., "solve this elliptic PDE with a relative accuracy of $10^{-6}$ and time less than 600 seconds") is provided to a recommender system that makes an initial recommendation of code modules (e.g., "use a finite-difference discretizer with red-black ordering"). These code modules are communicated to the compositional system as a "configuration", which are then scheduled and executed on the Grid; as the computation progresses (e.g., the PDE gets discretized and the resulting linear system appears to be ill-conditioned), feedback is provided to the runtime recommender through the checkpointing mechanism, which uses this information to perhaps dynamically insert a preconditioner before the linear solver in the solution loop. The configuration is updated with this selection, and the computation is re-scheduled (this time, perhaps migrated over to a different cluster on the Grid). This interplay between the compositional system (which supports object-based composition, migration, and checkpointing) and the runtime recommender (which enables dynamic selection of code modules) leads to a novel runtime framework for grid computations.

---

[1] Simplistically, the probability of no failures = $p^{200}$ (in our case), where $p$ is the probability of a node being operational. This leaves a large residual probability of failure.

It is pertinent to note that our solution approach supports all aspects of application composition over the Grid – **model specification, model execution, and model analysis**. Model specification deals with how representations for different aspects of a computation are brought together to create a representation of the computation as a whole [Forbus, 1996]. It addresses the ease of specifying Grid computations by the end-user. Model execution addresses the facility by which such scientific codes (constructed compositionally) can be scheduled, executed, and deployed over the Grid. Model analysis encompasses the ways by which performance information from execution is used to evaluate models and compositions, including supporting the refining and improvement of models. Issues of checkpointing, code instrumentation, and performance characterization are pertinent here. We use the term **compositional modeling** to collectively refer to all of the above three aspects, as they have become well accepted as an integral tenet of Grid computing.

## *In this Paper*

Section 2 identifies two core computational technologies that form the basis of our solution for runtime systems support. Section 3 elaborates on how these technologies are integrated to provide novel systems support for Grid computing. Section 4 identifies a set of "adaptivity schemas" that can be used as templates for realizing many complex, adaptive, scientific computations. Section 5 presents early results and outlines work in progress. A concluding discussion placing this work in context of current Grid computing is provided in Section 6.

# 2  Core Computational Technologies

Our approach to supporting adaptive compositional modeling on the Grid centers on two core computational technologies: the *Weaves* parallel compositional framework, and data-driven runtime recommender systems. We discuss them in detail in the context of a real scientific application.

## 2.1  Motivating Application

Our driver application involves the idea of *collaborating partial differential equation (PDE) solvers* [Drashansky et al., 1999] for solving heterogeneous multi-physics problems. For instance, simulating a gas turbine requires combining models for heat flows (throughout the engine), stresses (in the moving parts), fluid flows (for gases in the combustor), and combustion (in the engine cylinder). Each of these models can be described by an ODE/PDE with various formulations for the geometry, operator, and boundary conditions. The basic idea here is to replace the original multi-physics problem by a set of smaller simulation problems (on simple geometries) that need to be solved *simultaneously* while satisfying a set of interface conditions. The mathematical basis of this idea is the interface relaxation approach to support a network of interacting PDE solvers [McFaddin and Rice, 1992; see Fig. 1].

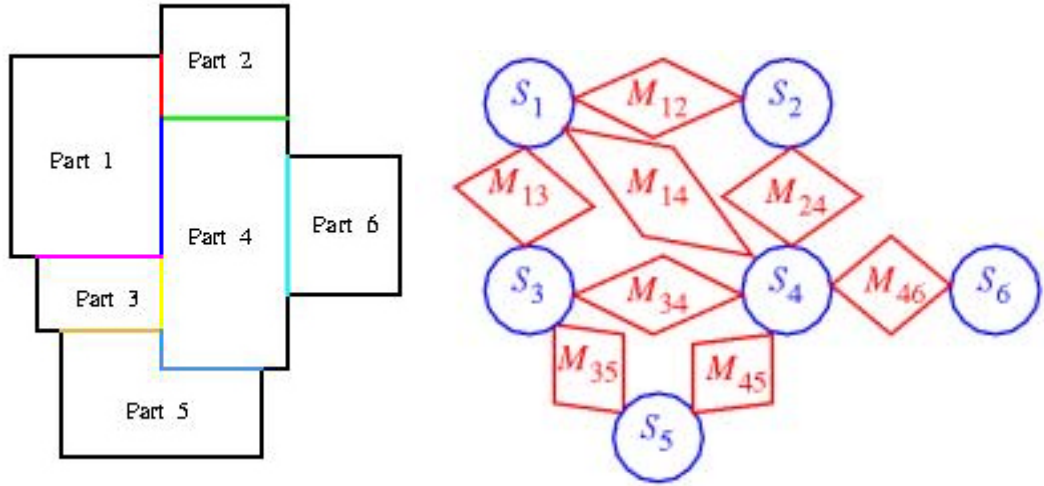

**Figure 1:** (left) Multi-physics problem with six subdomains with different PDEs. (right) A network of collaborating solvers (S) and mediators (M) to solve the PDE problem. Each mediator is responsible for agreement along one of the interfaces (colored lines).

Mathematical modeling of the multi-physics problem distinguishes between *solvers* and *mediators*. A PDE solver is instantiated for each of the simpler simulation problems and a mediator is instantiated for every interface to facilitate collaboration between the solvers. The mediators are responsible for ensuring that the solutions (obtained from the solvers) match properly at the interfaces. The term "match properly" is defined by the physics - if the interface is where the physics changes - or is defined mathematically (e.g., the solutions should join smoothly at the interface and have continuous derivatives). Distinguishing between solvers and mediators allows us to handle mathematical models naturally and elegantly; further, they can be organized to reflect the hierarchy of the physical structures (in this case, the turbine) underlying the computation.

In large-scale multi-physics simulations, it is not uncommon to have problems requiring collaboration between hundreds of solvers; one solver is assigned to each subdomain and the mediators issue instructions on appropriate boundary condition settings to their adjacent solvers. Once such a "network" of solvers and mediators is configured, it is scheduled for computation over the Grid. After every iteration, the mediators might proceed to "adjust" the boundary condition settings to ensure a better matching of solutions or, if the change of boundary conditions is smaller than the tolerance, might report convergence.

The need for this application to run effectively on different subsets of a grid - with varied and multilevel memory hierarchies - is the primary motivation for the research described in this paper. To simplify the discussion and to set the stage for describing the Weaves framework in the next section, we abstract the essence of the collaborating PDE solvers application into a relatively simple data-sharing problem. This problem arises when trying to exploit multiple levels of parallelism in PDE solver codes. We believe it is a common problem facing computational scientists trying to develop performance-portable codes for the grid.

Suppose we have several instances of a *solver* task $S_i$ running in parallel. These solvers need not be the same, e.g., in Figure 1, suppose $S_1$, $S_2$, $S_3$ and $S_4$ are instances of one solver (e.g., a standard finite difference method with direct Gaussian elimination), $S_5$ and $S_6$ are instances of another solver (e.g., one that uses the GMRES iterative method and a suitable preconditioner), and so on. These solvers are contributing to a shared state maintained by the *mediator* task $M_{ij}$. The challenge is to implement codes exhibiting these characteristics - independent tasks, organized at multiple levels of parallelism, sharing state amongst themselves at different levels - and to do so *transparently*. We argue in Section 2.2 that standard programming models (processes, threads) are not the complete answer. Our emphasis hence is on a compositional framework that can transparently support arbitrary state sharing for scientific computations. The *Weaves* parallel compositional framework embodies our solution approach to this problem.

The composite PDE solvers application also helps motivate the need for adaptivity in Grid computations; there are, literally, hundreds of well-defined software modules for supporting various aspects of the simulation process. There are multiple alternatives for numerical methods (iterative or direct solvers), numerical models (standard finite differences, collocation with cubic elements, Galerkin with linear elements, rectangular grids, triangular meshes), and various physical model assumptions and simplifications (e.g., cylindrical symmetry, steady state, rigid body mechanics, full 3D time-dependent physics). In addition, there are a variety of interface-relaxation methods [Rice et al., 1999] that can be implemented by a mediator. Performance information gathered at runtime can be fruitfully used to steer the dynamic selection of a suitable software module, which must then be linked in at runtime, executed, and possibly used to close the loop, to guide future compositions. Data-driven runtime recommender systems help organize the cataloging and mining of performance data for realizing adaptivity in compositional modeling.

## 2.2 The Weaves Parallel Compositional Framework

Weaves is a source-language independent parallel framework for object-based composition of *unmodified* scientific codes. Weaves works through reverse-compiler analysis; by analyzing compiled ELF object files, Weaves enables the vast repository of legacy scientific libraries to be seamlessly used in a object-based compositional framework, *without requiring that these codes be written in an object oriented language*. Formally, Weaves

- provides a source language independent framework based on object code analysis
- provides transparent checkpointing/recovery support. Object code analysis automatically determines the state that needs to be checkpointed and/or restored *without user intervention*.
- provides support for performance data gathering via code instrumentation.
- supports notions of both *spatial* and *temporal* adaptivity (defined later), a critical element of runtime compositional modeling
- supports runtime migration of fine-grain code modules. As opposed to process migration, Weaves allows the **migration of parts of a composed application** over the Grid.

As a compositional framework, perhaps the most important feature of Weaves is the modeling perspective it brings to bear on scientific computations; this enables scientific codes to be viewed in the context of a framework that integrates execution, simulation, and modeling of Grid applications.

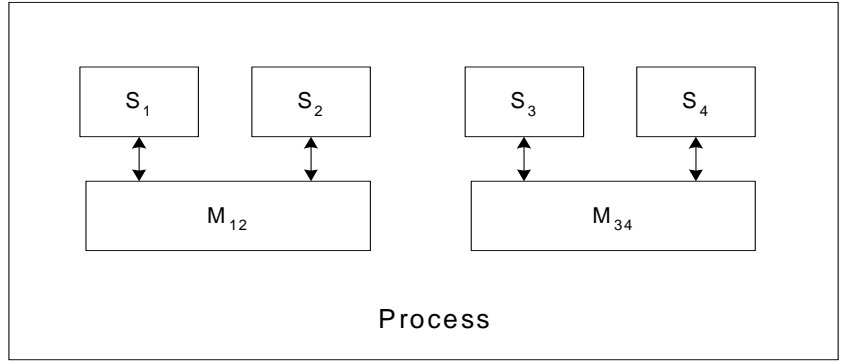

**Figure 2**: Design goal of the Weaves framework. We need to support multiple solvers linked to mediators within a single process.

Let us revisit the collaborating PDE solvers application from the context of the threads and processes models. We will illustrate the need for Weaves by a series of examples culminating in a compositional framework that can support the model shown in Figure 2. For simplicity, assume that $S_1$ and $S_2$ are two instantiations of the PDE algorithm $A_1$ and $S_3$ and $S_4$ are two instantiations of a different PDE algorithm $(A_2)$.[2] From a scalability perspective, the compositional framework should operate within a single process, *with multiple concurrent flows of control*.

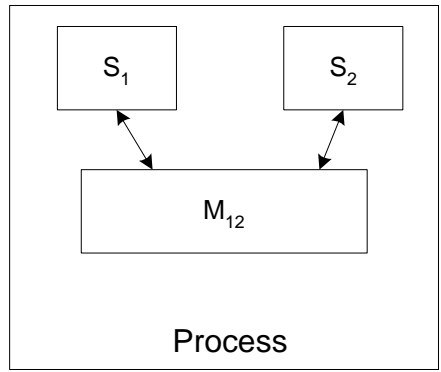

**Figure 3:** A simple compositional model with two PDE solvers linked to a single mediator. This composition addresses a composite PDE problem involving two domains.

Let us start with a simple composition depicted in Figure 3 where two PDE solvers are linked to a single mediator performing an interface relaxation. To model this interaction, we use a simple *process per solver* model as shown in Figure 4a. This model allows a single PDE solver application to be linked to a mediator. The external references of the PDE solver application will be bound to the mediator. Multiple such processes can be executed to simulate a network of collaborating solvers. This simple approach has several problems. The first and most basic problem is that we cannot link multiple solver applications to a *single shared* mediator without modifying the application. The second major problem with the process model arises from scalability concerns. A large network of collaborating PDE solvers will involve tens to hundreds of interfaces, each of which is a process. Inter-process context switch time will be a major bottleneck.

---

[2] This is not a requirement of the Weaves framework. Technically, all the solvers could implement different algorithms. However, the interesting cases arise when there are multiple instantiations of a given PDE algorithm.

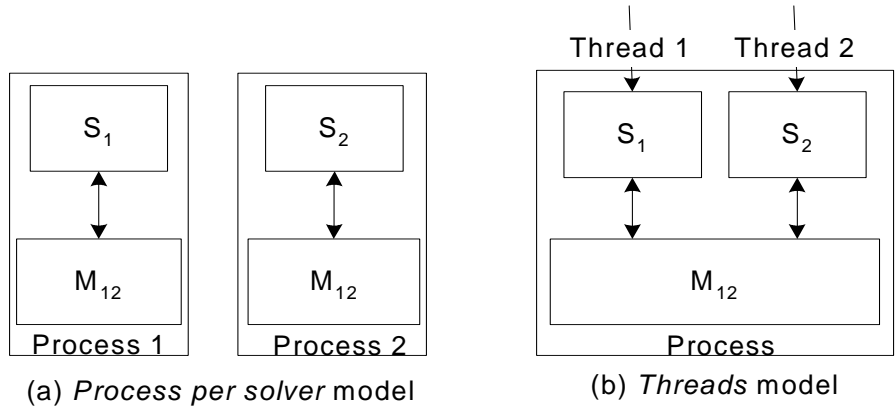

(a) *Process per solver* model     (b) *Threads* model

**Figure 4**: The composition shown in Figure 3 modeled under (a) process per solver model and (b) threads model. Neither of these models can achieve the composition shown in Figure 3.

To address the above concerns, let us build a new model based on threads as shown in Figure 4b. In the threads model, the composition shown in Figure 3 can be achieved by modifying the PDE solver application source to create two threads, or inserting a piece of stub code that creates two threads, with the start function of each thread set to the entry point of the solver application. As before, the mediator application is linked to the PDE solver.

The major problem with this approach arises from updates to global variables - in particular, the PDE solver may not be thread-safe. Since threads share global variables, a solver thread modifying a global variable will inadvertently change the state of the other – unrelated - solver thread causing erroneous behavior. Ideally, we need two copies of all the global variables used in the PDE solver. In programs that are explicitly threaded in design, sharing of global state is intentional. In our case, this *sharing is neither intentional nor necessarily desirable*. On the other hand, to create a shared mediator, we need to share global state within the mediator between the threads running through it.

The threads example illustrates the crux of our problem – conflicting needs – the need to **avoid sharing** global state between threads of the solver application and the need **for sharing** global state between the threads running through the mediator. The intuition here is that we need a programming model that allows arbitrary sharing of global state. Such a model can subsume both the thread and process models, since it can allow both complete sharing of global state as in threads as well as no sharing as in processes This observation leads us to the first step towards Weaves compositional framework.

As an aside, we mention that recombination of global state can be achieved through an agents model. It is thus not surprising that the collaborating PDE solvers application has been approached using agents technology [Drashansky et al., 1999]. The critical observation here is that messages in agent technology are a powerful code-neutral abstraction for parameter passing and procedure invocation. Effectively, messages between agents are used to recombine state. For instance, two solver processes are used to separate their global state. To recombine state, the solver processes communicate with a single mediator process, whose state is a function of the messages received from the solvers.

### 2.2.1 Defining the Weaves Framework
The major components of the Weaves programming framework are:

- **Module**: A module is any object file or collection of object files defined by the user. Modules have:
    - A **data context**, which is the global state of the module scoped within the object files of the module, and
    - A **code context**, which is the code contained within the object files that constitute the module. The code context may have multiple entry point and exit point functions.
- **Bead**: A bead is an instantiation of a module. Multiple instantiations of a module have independent data contexts, but share the same code context.
- **Weave**: A weave is a collection of data contexts belonging to beads of different modules. The definition of a weave forms the core of the Weaves framework. Traditionally, a process has a single name space mapped to a single address space. Weaves allow users to define multiple namespaces within a single address space, with user-defined control over the creation of a namespace.
- **String**: A string is a thread of execution that operates within a single weave. Similar to the threads model, multiple strings may execute within a single weave. However, a single string cannot operate under multiple weaves. Intuitively, a string operates within a single namespace. Allowing a string to operate under multiple namespaces would violate the single valued nature of atomic variables.
- **Tapestry**: A tapestry is a set of weaves, which describes the structure of the composed application. The physical manifestation of a tapestry is a single process.

The above definitions have equivalents in object-oriented programming. A module is similar to a class and a bead - which is an instantiation of a module – is similar to an object. Tapestries are somewhat similar to object hierarchies. The major exception is that interaction between beads within a tapestry involves runtime binding. We chose to use our own terminology to (i) avoid overloading the semantics of well-known OOP terms and (ii) avert the implication that the framework requires the use of an OOP language.

Strings are similar to threads in that (i) they can be dynamically instantiated and (ii) they share the same copy of code. However, unlike threads, strings do not share global state. Each string has its own copy of global state. The main goal here is to avoid inadvertent sharing of state between unrelated instantiations of an algorithm, without having to modify the algorithm.

Since strings are an intra-process mechanism, we will illustrate their operation by comparing and contrasting them to threads. A thread's state consists of (i) an instruction pointer (**IP**), (ii) a stack pointer and (iii) copy of CPU registers. Each thread within a process has its own stack frame that maintains local variables and a series of activation records that describes the execution path traversed by the thread. When a thread is created, the thread library creates a new stack frame and starts execution at the first instruction of the function specified by the thread instantiation call. When the thread scheduler needs to switch between threads, it saves the current IP, current stack frame, and the values in the CPU registers, switches to the state of the next thread, and starts execution at the IP contained in the thread state.

Strings involve an extension to the operation of threads. Similar to threads, each string has its own stack frame, which maintains local state. In addition, each string also has a copy of the global variables in an area called the ***weave context frame***, the start of which is pointed to by a ***weave context frame pointer***. A weave context defines the namespace of a string. This includes the global variables of all the beads traversed by a string. Note that some of the beads in a string may be shared between strings.

A string's state consists of (i) an instruction pointer, (ii) a stack frame pointer, (iii) copy of CPU registers, and (iv) a weave context frame pointer. When a string is created, the string creation call creates a stack frame and a weave context frame (if necessary) and copies the current state of the global variables into the weave context frame. The string creation call also associates a numerical identifier with the newly created string. Since creating a string involves copying its global variables, the string creation cost depends on the storage size of the global variables resulting in a higher creation cost than threads. We justify this cost by noting that it is a one time cost paid at program startup. Also, well-written applications are generally frugal in their use of global state, which mitigates the impact of the copy operation.

Similar to a thread scheduler, the string scheduler starts execution of the new string at the first instruction of a user-specified function. When the string library needs to switch between strings, it saves the current IP, current stack frame pointer, the values in the CPU registers, and the current weave context frame pointer, switches to the state of the next string and starts execution at the IP contained in the string state. The inter-string context switch cost is identical to threads.

Selective sharing of state in our framework operates at the level of individual beads. We illustrate the operation of selective sharing with the example shown in Figure 2 (also repeated in the Figure 5 below). The tapestry defines 4 weaves <Solver $S_1$, Mediator $M_{12}$>, <Solver $S_2$, Mediator $M_{12}$>, <Solver $S_3$, Mediator $M_{34}$> and <Solver $S_4$, Mediator $M_{34}$>, and 4 strings, with each string operating within a single weave. At run time, context switching between the strings automatically switches the namespace associated with the string, preserving the sharing specified in the tapestry.

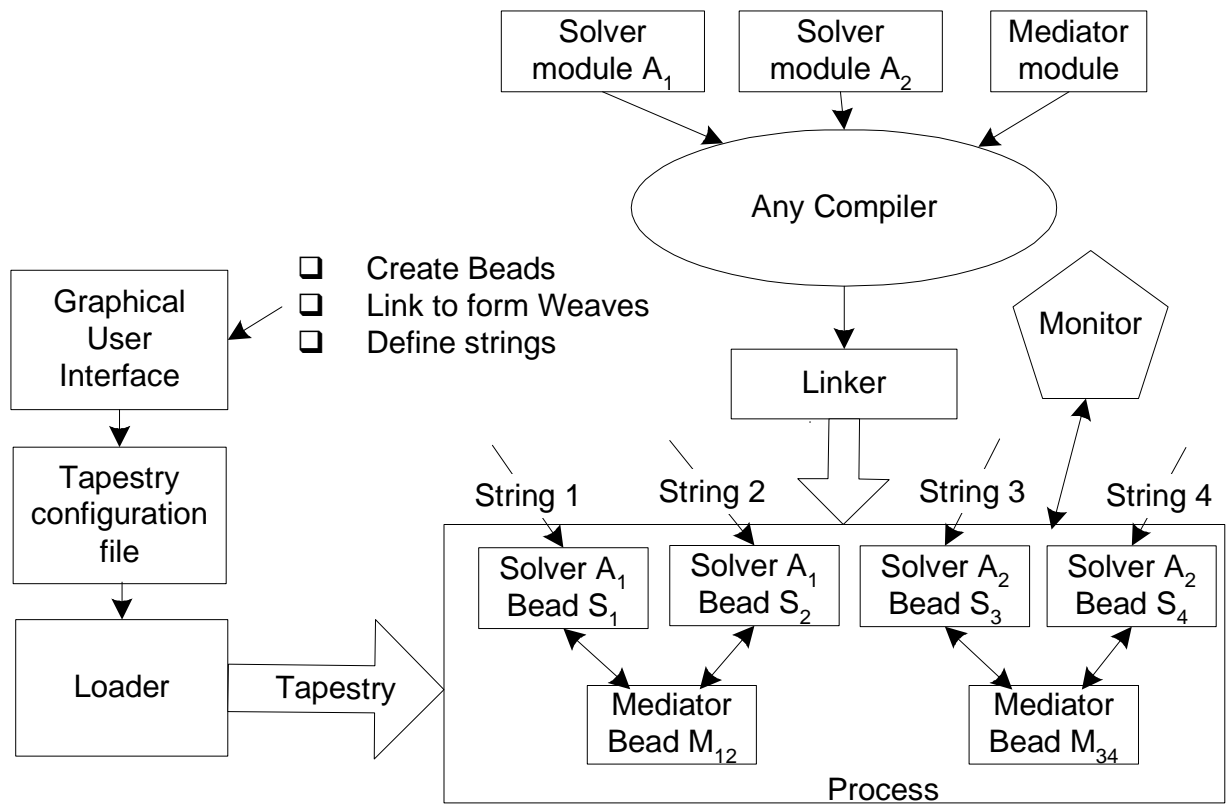

**Figure 5:** Interaction between the various components of the Weaves framework.

Figure 5 depicts the design process in the Weaves framework. The design process involves two entities: a *programmer* who implements the modules and a *composer*, who uses a graphical user interface to instantiate beads and define the various weaves and strings. The result of the GUI composition is a tapestry configuration file, which is used to load and execute the composed application. Each composed application also has a module called a *monitor* that is automatically linked with the composed application. In the process model, utilities like *ps* (in UNIX) can be used to query the run time of the process. The monitor provides a much more powerful IPC (Inter Process Communication) interface to such functionality. Utilities can query the monitor to determine the current tapestry, beads, strings, and weaves within a composed application.

### 2.2.2 Runtime Reconfigurability

The tapestry generated by the GUI is not necessarily a static composition. The Weaves programming framework allows applications to rewire themselves on the fly in response to dynamic conditions. Two forms of dynamic application composition are supported in the framework. In the first form, if the requisite modules are already linked into the original tapestry, Weave-aware applications can modify their structure by creating new beads, defining weaves, and instantiating strings at run-time. For non-Weave aware applications, the interface exposed by the monitor can be used to modify the tapestry of a composed application. These modifications may be manually made by a user at the command line or can be automatically generated by an external *resource monitoring* agent.

In the second form of dynamic composition, new code modules can be inserted into a running application through a modified dynamic library interface. In this mode of operation, the dynamically inserted code is analyzed at run-time. Dynamically inserted modules can be used in the same manner as statically inserted modules. This interface provides the full capabilities of Weaves, including arbitrary namespaces and compositional capabilities, in a run-time compositional framework. We will exploit this capability to investigate runtime algorithm selection and composition (see Section 3.1).

### 2.2.3 Scheduling Non-Reentrant Codes

Since the Weaves framework does not require that codes be re-entrant (thread-safe), scheduling poses several interesting problems. When an operating system or threads scheduler preempts a task, it can switch the operating context to any other task that is ready to run. In contrast, in the Weaves framework, preemptive scheduling strategies can cause reentrancy in beads shared at lower layers. If the codes are not reentrant, this will result in incorrect operation.

We solve this problem by organizing strings into equivalence classes, where each equivalence class contains strings that share beads. Preemptive scheduling switches between strings of different equivalence classes. If the preempted string has not traversed a shared bead, preemptive scheduling can also switch between strings of the same equivalence class. In effect, the scheduler makes constant-time cuts on the execution of a weaved application to determine if it is "safe" to switch the string context. While this solution ensures that the weaved application as a whole is making progress, it does not guarantee against starvation of strings belonging to the same equivalence class.

To guarantee against starvation of strings belonging to the same equivalence class, we insert *continuations* for entry/exit point functions in shared modules. The continuations *cooperatively* relinquish string control *immediately* after control returns from traversing a shared bead. A combination of such cooperative scheduling with traditional preemptive scheduling ensures that as long as there are no inherent starvation and/or deadlock conditions in the original code, the Weaves framework will not introduce string starvation. The concluding section contains a discussion on automatic deadlock detection and recovery for Grid applications.

### 2.2.4 Tuple Spaces

The notion of selective state sharing in the Weaves programming framework presents a very powerful mechanism for defining namespaces. Since the definition of a weave permits any set of beads to define a namespace, any composition that can be represented by a connected graph (or a set of independent graphs) can be realized by this framework. From an application's perspective, the definition and operation of distinct namespaces is transparent. This mechanism presents a powerful compositional framework for any procedural code.

The Weaves framework also supports the notion of shared tuple spaces. In the current definitions, distinct beads of the same module have different data contexts, i.e., data sharing occurs at the granularity of an entire module. To create a shared tuple space, we need fine grain control over the individual members of a data context.

In order to support shared tuple spaces, from the perspective of the framework, we need mechanisms to (i) define a shared tuple space and (ii) to selectively share the members of the tuple space across multiple beads. To define a shared tuple space, application composers can use the graphical user interface to denote the members of the tuple space or code modules can use a syntactic notation to mark the members of the tuple space. This information is used at bead creation time to merge references to shared members of a tuple space.

### 2.2.5 Automatic Checkpointing and Recovery

A primary goal of the Weaves framework is to support adaptive applications that can rewire themselves dynamically in response to changing conditions. In the parallel discrete event simulation context, our view of adaptivity encompasses optimistic algorithms that try to take the best execution path given a set of available options. However, the path chosen may not always be right, requiring applications to rollback to a known correct state. As discussed in the introduction, typical HPC applications also require checkpointing and recovery.

Traditionally, state checkpointing and restoration has been left to individual applications. This significantly adds to the complexity and maintainability of such codes. Furthermore, event driven codes add an additional layer of complexity. Since the path of execution through an event driven application is not known apriori, checkpointing and restoring such applications present significant challenges. Our goal here is to provide a *transparent support* framework that can checkpoint and recover state, *without application support*.

To provide support for automatic checkpointing and recovery, note that in the Weaves framework, each string maintains its global variables in the weave context frame and local variables and call invocation history in the stack frame. This compartmentalizes static state into two well defined regions. We can save the contents of the stack and weave context frames, effectively saving static state. However, this does not account for dynamic memory allocated during runtime.

To track dynamic memory allocation, we use a mechanism similar to the one used by memory leak debuggers. We overload the library calls responsible for dynamic memory allocation – *malloc()*, *calloc()*, *realloc()*, and *free()* in C. The overloaded calls keep track of the bead identifier, the start of the memory region and the size of allocated memory.

We now have access to both the static as well as dynamic state of the tapestry, which can be used to implement checkpointing and recovery. The naive mechanism for checkpointing involves (i) saving the contents of the stack frame, (ii) saving the contents of the weave context frame, and (iii) copying the contents of all dynamically allocated memory regions. Restoring application state involves garbage collection of all dynamic memory allocated after the checkpoint and restoring the state saved during the checkpoint. It is easy to see that the naive approach is not memory efficient, particularly in our domain where tapestries can contain hundreds of beads.

To implement an efficient checkpointing mechanism, note that operating systems already have efficient mechanisms for handling process fork calls, through the use of *copy-on-write* semantics. A sophisticated approach to checkpointing can be implemented with the *mprotect* POSIX system call, which implements a new light weight version of the copy-on-write mechanism that operates in an intra-process domain. When a checkpoint is invoked, we mark all data pages corresponding to dynamically allocated memory and the weave context frame read-only. As the application proceeds, updates to read-only data cause segmentation faults (SIGSEGV), which are handled by duplicating the offending page and allowing read/write operation on the duplicate. This mechanism works optimistically, limiting the memory overhead of checkpointing to only modified data.

### 2.2.6 Weaves in a Grid Environment

The future of Grid computing hinges on its ability to provide a seamless view of distributed computational resources. Facilitating this view requires us to reconcile the starvation concerns of large computational applications that run across distributed resources, with the administrative control needs of the smaller individual units that comprise the Grid. This leads to one of the most challenging cases for automatic checkpointing, recovery, and migration – namely, that of large-scale, distributed-memory, message-passing codes. While this problem is motivated by reliability concerns for parallel codes running on large clusters, it is especially critical for grid computing codes, where checkpointing and runtime migration are essential to enable fluidic control over administrative resources. In the last two months, we have developed a user-level implementation of the TCP/IP stack (as a part of a project on scalable network emulation) which, combined with the Weaves framework, provides a promising new approach to checkpointing and runtime migration of parallel codes.

There are several serious issues with checkpointing parallel codes written using MPICH (a popular implementation of the MPI library) or PVM. First, we need a reliable single process checkpointing tool; these are not available on all platforms. While a single process checkpointing tool can capture process state, we also need mechanisms to capture state maintained within the operating system on behalf of the process. This includes open network socket handles and network data within operating system buffers. One approach to this problem is to create a consistent global state of the parallel application [Chandy and Lamport, 1985] and then initiate the checkpoint operation. While this approach works, it still doesn't enable migration. MPI implementations maintain significant static environment state within themselves, including IP addresses, hostnames and open TCP connections. To migrate a checkpointed MPI application, we need mechanisms to update the internal MPI state to reflect the change in the underlying environment, which makes any implementation of such a system specific to a particular MPI or PVM codebase.

Our goal is to implement a transparent checkpoint and migration framework for *any* parallel communication library that uses the TCP/IP protocol stack. The novel idea here is to use partial consistency instead of global consistency to derive the unified state of the application. In our notion of partial consistency, we do *not* checkpoint state within the operating system or in-flight over the communication fabric. Instead, we exploit the ability of a reliable communications protocol to mask this loss of state. The main advantage of this approach is that enables checkpointing at *any* point in the execution of a parallel application, including within barrier operations.

In our design, a parallel application is linked to a message passing library (e.g., MPI), which in turn is linked to our TCP/IP implementation through a standard sockets interface. Our TCP/IP implementation treats the underlying communication subsystem as an unreliable link and only requires support for two calls, a transmit call and an inbound receive call. Another interesting feature here is that the implementation of reliability in our TCP/IP stack is decoupled from real-time by using *interval timing*, where the interval durations are preserved across checkpoint/restart procedures.

The main advantage of this design is that the user-level TCP/IP stack provides an additional layer of indirection, exposing a common IP address and hostname independent of the physical platform or underlying communication infrastructure. We propose to use the Weaves framework to deliver platform independent checkpointing and migration facilities. The Weaves framework will checkpoint the entire application, including the user-level TCP/IP implementation. When a parallel application is migrated, the user-level TCP/IP implementation is used to abstract the specifics of the target environment from the application. There are several additional advantages to this approach. First, a parallel application can be moved to a platform with a different underlying communication infrastructure. Secondly, the same framework can be used to simulate different physical parallel communication fabrics and analyze the performance of communication libraries. Finally, this approach enables easy portability of parallel communication libraries, since they can be developed for TCP/IP and rely on our system to provide the actual mapping to the physical environment.

## 2.3 Runtime Recommender Systems

Recommender systems [Ramakrishnan et al., 1998] provide facilities for automatic knowledge-based selection of solution components on the Grid. They help make selections of algorithms and code modules by taking into account both problem characteristics and performance considerations. Recommender systems involve the empirical evaluation algorithms on realistic, often parameterized, test problems, and interpreting and generalizing the results to guide selection of appropriate mathematical software. They are the preferred method of analysis in applications where domain knowledge is imperfect and for which our understanding of the factors influencing algorithm applicability is incomplete. For instance, when solving linear systems associated with finite-difference discretization of elliptic PDEs, there is little mathematical theory to guide a choice between, say, a direct solver and an iterative Krylov solver plus preconditioner. A recommender systems approach is to parameterize a suitable family of problems, and mine a database of thousands of PDE "solves" over the Grid to gain insight into the likely relative performance of these two approaches (e.g., see Figure 6). Parameter sweep templates for Grid computing [Casanova and Berman, 2002] are thus an important tool for designing recommender systems.

In a traditional design of a recommender system [Ramakrishnan and Ribbens, 2000; Houstis et al., 2000], a database of test problems and algorithms is organized, and performance data is accumulated for the given problem population. This database of performance data is then mined (generalized) to arrive at high-level rules that can form the basis for a recommendation (for future problems). A variety of data mining algorithms are appropriate here (e.g., attribute-value generalizations, inductive logic programming; see [Houstis et al., 2000]). The MyPythia Grid portal [Houstis et al., 2002] provides many interfaces to these algorithms, for both the recommender system builder and the recommender system user. In this system, the data collection phase is distinct from the generalization aspect (we refer to these as "offline" recommender systems); in other applications, data collection occurs in conjunction with data mining [Ramakrishnan et al., 2002], so that it can be "steered" to more accurately sample desired regions of the recommendation space.

In a Grid setting, recommender systems are important aids to application composition, by making dynamic selections of components (we refer to these as "runtime" recommender systems). Such a facility is important in many problem domains because: (i) the nature of the problem being solved changes as the computations are being performed, (ii) the underlying computing platform or resource availability is dynamic, or (iii) information about application performance characteristics is acquired at runtime, from the actual computation rather than from offline analysis.

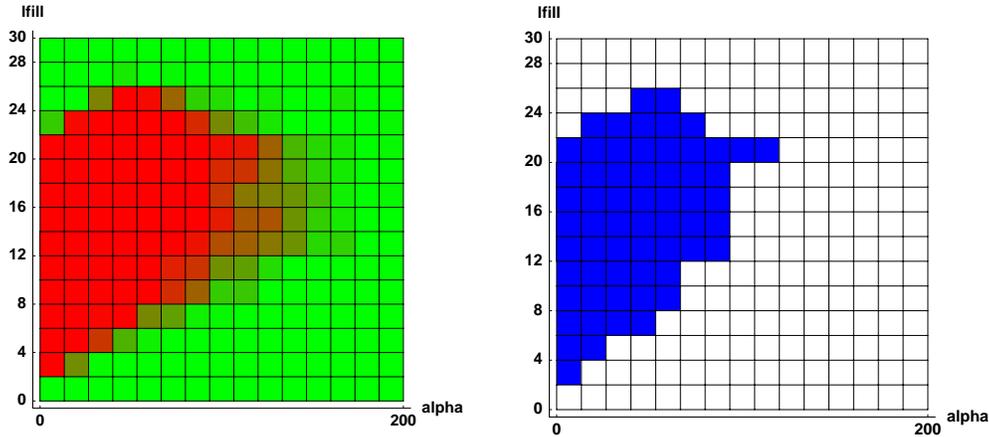

**Figure 6:** (left) Mining and visualizing recommendation spaces for selecting between a GMRES iterative solver (red) and a direct Gaussian elimination solver (green) to solve an elliptic PDE. *α* is a parameter controlling the singularity in the PDE problem (and hence, the ill-conditioning of the corresponding linear system) and *lfill* controls the pre-conditioning in the iterative solver. (right) A mined recommendation space with 90% confidence, showing the region where the GMRES solver is preferred. As *α* grows larger, it is seen that the *lfill* parameter must fall within a narrower range for the iterative solver to be preferred, until eventually the direct solver becomes the preferred choice. For more details, please see [Ramakrishnan and Ribbens, 2000].

The importance of a runtime recommender is easily seen in applications such as the collaborating PDE solvers, where selections need to be made of a discretizer, preconditioner, and linear system solver (in that order). Information needed to make a preconditioning recommendation or linear solver recommendation is not available *until after* the PDE has been discretized, hence such recommendations have to happen at runtime, using dynamic information. Specifically, a runtime recommender monitors a computational process, detects state-changes, and makes selections of solution components dynamically, thus aiding knowledge-based application composition at runtime. Designing a runtime recommender is thus more involved than an offline recommender because the database of problems and algorithm executions is not readily available and needs to be captured "on the fly".

### 2.3.1 Strategies for Runtime Recommendation

The primary problem faced by a runtime recommender is to observe a computational process (as it unfolds), make recommendations along the way, with the added complexity that feedback (about recommendations) is not immediate, and will arrive several timesteps (typically unknown) later. This is a problem reminiscent of *reinforcement learning* [Kaelbling et al., 1996][Sutton and Barto, 1998], well studied in the control systems and AI literature (and was one of the main influences in bringing control-theoretic techniques to realize adaptivity in scientific software; e.g., see [Gustafsson, 1991]). Note that the task here is more ambitious than mere parameter tuning or building expert systems. The key issue is to tradeoff the cost of exploring the environment in the short-term with an accuracy improvement in the long-term. A runtime recommender systems thus grapples with a constant dilemma: should it choose a solver that it knows has worked before (**exploitation**) or should it "try" a different solver to see if it might lead to a performance improvement (**exploration**)?

Our approach to this problem is to model the scientific application as a non-deterministic, stationary system (the transition probabilities between states are assumed to be constant to ensure convergence of the learning algorithms). This network does not need to be handcrafted, but can be constructed online by the recommender system. For example, in the PDE application, "states" correspond to physical stages of the computational process and are represented by features such as singularity, current algorithm, order of the method, and performance criteria (set by the user). The "actions" correspond to choices made by the recommender, such as "use the ILU preconditioner", "switch from iterative to direct method", "decrease the current order". The goal now is to learn the *utility* of taking certain actions in various states. These utility estimations are summarized in the form of a control policy that chooses the action with the highest utility. On each step of the interaction, the recommender receives as input some indication of the current state (such as problem features) and it generates a recommendation as output. This recommendation changes the state of the system (e.g., at the end of the first stage of PDE solution, information about linear system characteristics becomes available), and the value of this state transition is communicated back to the recommender as reinforcement; which then chooses recommendations that will tend to increase the long-run sum of values of the reinforcement signal. Once again, there are a variety of learning algorithms for iterative improvement of generalizations.

The recommender begins in a mode that favors exploration over exploitation. These runs are typically scheduled during idle cycles on our computational grid. Over time, the recommender encounters enough problems from its database and has explored enough alternatives that it can make an informed judgement about solution alternatives. At this point, its mode of operation becomes primary exploitative with only a small percentage of exploitation (to ensure that the learned utility values are current). Such an approach has been validated for selecting quadrature routines from a space of over 120 algorithms [Ramakrishnan et al., 2002] and for synthesizing type-insensitive codes for ODEs with both stiff and non-stiff regions (see Figure 7 for a policy mined by inductive logic programming). The functionality provided by a runtime recommender can be thought of as automatic determination of control policies to realize adaptivity in scientific codes. Runtime recommendation is traditionally concerned with code executions but can also be employed to assess model-based simulations and make selections of system configurations, as studied in the adaptive control formulation of [Adve et al., 2002].

```
qvalue(1) :-
    state(beginning), algorithm(none),
    action(choose-non-stiff).

qvalue(1) :-
    state(near-stiff), algorithm(non-stiff),
    estimate < threshold, sv < 10,
    action(switch-to-stiff).
```

**Figure 7:** A partially induced control policy mined by a runtime recommender for the task of solving ODEs with both stiff and non-stiff regions. From the beginning state, the recommender always prefers a non-stiff method (an Adams-Moulton method), but when its estimates improve its assessment of the evolution of solution components, it switches to a stiff method (in this case, an implicit A-stable formula). This approach should be contrasted to classical differential equation software such as GEAR, LSODE, and DIFSUB where such adaptivity is realized through static decision-making code coupled with the ODE solver.

# 3   Systems Support for Adaptive Compositional Modeling

To summarize the two core technologies: runtime recommender systems allow the dynamic selection and composition of code modules and *Weaves* provides runtime systems support to realize such compositions. In this section, we further bring out the synergy between these technologies.

From the viewpoint of the Weaves design process, a runtime recommender system acts as the composer, dynamically determining the code modules and their instantiations. Specifically, the runtime recommender supplies the tapestry configuration file that specifies the composition graph. From the viewpoint of the recommender system, Weaves acts as the "end-effector" and provides systems support for checkpointing algorithm executions and modifying code modules in response to changing problem or platform characteristics. Figure 8 depicts how Weaves and the recommender system interact to provide an adaptive runtime framework for Grid computations. Recommendations are made available as a tapestry configuration file, which are then used to either create dynamic instantiations/code insertions or statically linked executables. The realized tapestry is then made available through multiple interfaces that guide its future execution. Note that recommendations happen in the context of a single process (which may have multiple concurrent control flows). In the next section we show how parts of recommended compositions can be migrated and scheduled on Grid infrastructures. As Figure 8 shows, the integrated framework provides transparent application-level resource management and fault tolerance capabilities. Grid services such as Globus can use these interfaces to implement fault recovery and migration in the context of the entire Grid infrastructure.

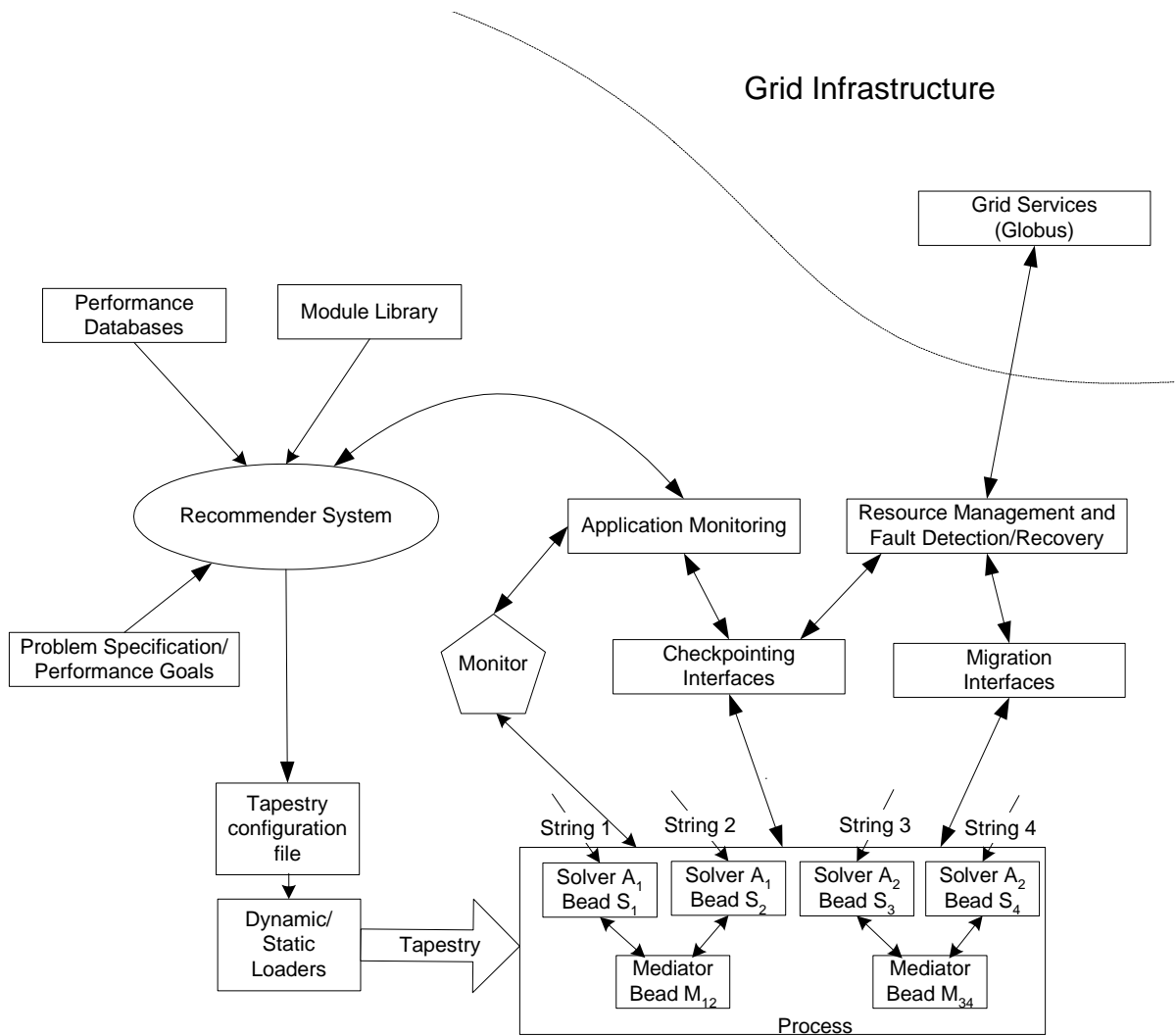

**Figure 8:** Interaction between the runtime recommender system and Weaves in the context of Grid infrastructures.

## *3.1 Temporal and Spatial Adaptivity*

Consider how the interaction between Weaves and runtime recommender systems would work for the task of adaptive numerical quadrature. Let us start with the collection of 120 quadrature algorithms described in [Ramakrishnan et al., 2002]. For a given numerical integration problem, the performance goal is to recommend a suitable quadrature routine such that the number of function evaluations is minimized. A recommender system (GAUSS) with this functionality is also described in [Ramakrishnan et al., 2002]. GAUSS can make suitable selections of algorithms from the quadrature library, monitor their execution, and change its recommendation if its earlier selection failed or otherwise did not satisfy the performance constraints. GAUSS is more than a polyalgorithm comprising of the 120 algorithms (where the decision procedures for selection are hardwired); it has the ability to use runtime information about algorithm performance dynamically as it becomes available. It functions by organizing a database of parameterized test problems and algorithm executions and uses an online mechanism to continually generalize from archived performance data.

This mode of operation in GAUSS can be viewed as a form of *temporal adaptivity*. The Weaves framework provides two notions of temporal adaptivity. In the first form – called *pessimistic temporal adaptivity* – the recommender system can dynamically select an algorithm already compiled into the executing application (using a form of an if-then-else construct) but doesn't have the ability to "retract" its recommendation. For this form of adaptivity to be successful, the recommendation space must be limited in its choice of algorithms to *only* those that produce correct results. Pessimistic temporal adaptivity is thus most suited for the exploitation mode of a recommender system. In the second form – called *optimistic temporal adaptivity* – the recommender system can dynamically choose from *any* algorithm that suits the purpose, including those that *may not* produce correct results all the time. Optimistic temporal adaptivity is ideal for the exploration mode of a recommender system.

Optimistic temporal adaptivity requires runtime systems support for checkpointing and recovery, since we need to (i) recover from failed instantiations of algorithms, and (ii) ensure that the recovery process doesn't result in the recommender system following the "failed" path again. Note that (ii) is a rather insidious issue. A perfect checkpoint/recovery mechanism will restore the recommender system state to just before its selection of the failed algorithm, which will result in the recommender system following the "failed" execution path repeatedly. What is needed is a checkpointing and recovery system that can provide a tuple (any set of variables defined in the namespace of the executing algorithm) view of the future, where the tuple presents intermediate results. In addition to pruning the search process of the recommender system, the tuple may be used to augment the features gathered by the recommender, helping it make a more informed decision.

Optimistic temporal adaptivity is a very powerful mechanism for supporting runtime recommendation and composition. The tuple view of the future provides not just algorithm state (in the form of variables comprising the tuple), but also the entire function invocation history prior to the failure (which includes the entire sequence of algorithm recommendations exercised). For applications such as adaptive quadrature, which are based on a divide-and-conquer strategy of repeated problem decompositions and algorithm recommendations, this feature is particularly important.

Weaves supports a further form of adaptivity called *spatial adaptivity*, where even the space of algorithms to be selected for composition is not known until runtime. For instance, consider a molecular electronics simulation, where a sequence of thousands of linear systems have to be solved to compute the I-V profile of a single device. The complete realization of such a simulation may span weeks to months. During execution new solvers may become available - especially in grid settings - which may offer better performance characteristics. What we need is a mechanism to transparently substitute the solver compiled into the executing binary with a new solver "on-the-fly" – dynamic function replacement. This notion is similar to dynamic classloaders in Java™.[3] The Weaves framework provides strong support for spatial adaptivity, including across multiple non-OOP source languages.

## 3.2 Runtime String Migration

Weaves is inherently a parallel run-time compositional framework. The examples above show the operation of the framework within the context of a single process. In this mode of operation, the framework provides load balancing on shared memory multiprocessor architectures by executing different strings on different processors. This form of load balancing is largely transparent to the design of the application.

Since the current generation of large parallel supercomputers is based on distributed memory design, we need to extend the framework to provide similar *transparent* load balancing capabilities to distributed memory codes. Currently, distributed memory codes implement their own load balancing. Since we know the structure of the composed application, we are in a position to both expose additional interfaces as well as augment existing load balancing capabilities.

In large tapestries composed of tens of thousands of beads, good load balancing is necessary to obtain reasonable speedup. The scale of the system combined with incomplete knowledge of runtime load makes it nearly impossible to statically load balance such a system. Dynamic load balancing guided by runtime analysis of load is necessary to ensure scalability. Our view here is that load balancing in a distributed memory environment translates to run-time code migration. This will ensure that all participating host processors experience equal computation and communication loads. The main issue is determining the resolution of code migration. Should code migration occur at the level of individual beads or something larger?

To answer this question, let us take a look at the issues involved in code migration at different resolutions. In the case of an individual bead, we can track its static state but it is much harder to keep track of its dynamic memory allocation. To see why, let's take the example of a bead that invokes a function within another bead. The target function allocates an array of pointers, allocates memory to each element of the array and returns a pointer to the start of the pointer hierarchy. Our overloaded memory allocation calls will incorrectly attribute the memory allocated to each element of the array to the bead corresponding to the target function. Without exhaustive analysis and significant runtime support, it is not possible to track dynamic memory allocation within a bead.

To avoid the above problem, our observation is that while it may not be possible to track the dynamic memory usage of a single bead, it may be easier to track the total dynamic memory used by a set of beads. In essence, we are trying to create closed regions of interacting beads ─ *an island of beads* ─ that exchange memory between them, but have no connection to other beads. In general such islands of interacting beads can be found in most applications ─ they represent entities at a higher layer of abstraction. Graph theoretically, an island of beads represents a closed graph with no external connections. The modular framework of weaves also aids the process of isolating islands of beads by forcing designers to look at bead interactions during application composition. The GUI front-end used to create a tapestry can also be used to mark specific islands of weaves, which then become targets for code migration. Our target platform ─ a workstation cluster ─ uses the SPMD model of execution [Darema, 2001]. Since the same program executes on all cluster nodes, the necessary code modules either already exist at the target of the migration, or can be instantiated at run-time through the dynamic library interface. Code migration then reduces to the problem of instantiating new weaves at the target node corresponding to a newly migrated island and migrating the state corresponding to the island. Since weaved code always uses indexed addressing, the migrated code does not need any code patching to be functional.

---

[3] Java implements dynamic classloaders through its VM. Compiled OOP languages such as C++ can do late binding of function calls at runtime, but the target of the call has to be compiled into the executing binary. Weaves supports source-language independent late binding, including cases where the target of the call is dynamically loaded and linked.

The above analysis ignores a very serious problem with intra-process code migration – *pointer aliasing*. Traditionally code migration has been handled at the process level, resulting in process migration. Since each process operates within its own address space, when a process is migrated, it sees the same virtual memory addresses on the target processor. In our domain, migration occurs at the level of string, which is an intra-process entity. When a string is migrated, it will not necessarily get the same virtual memory addresses on the target processor. Code patching will be needed to fix the addresses. This significantly complicates migration in distributed memory machines.

This problem is even more complex than the above description. To see why, let us take a case where a bead allocates dynamic memory to a pointer variable *ptr*. It then sets a second pointer variable *ptr1* to *ptr*, i.e. *ptr1* points to the same memory location as *ptr*. When we migrate this code, we allocate dynamic memory at the target machine and set *ptr* to point to this memory. However, *ptr1* is still pointing to the old memory location from the source processor. There is no guarantee that the same memory location in virtual memory address region is available on the target processor. Not only do we have to fix memory addresses allocated dynamically, we also need to ensure that all aliases of memory addresses are fixed appropriately, a problem known as pointer aliasing.

Pointer aliasing is a significant research issue. Current solutions are based on restricting source language semantics to prevent pointer aliasing, or executing code within virtual machines. Neither of these solutions is available to us. We have no control over the source language and emulating code over a virtual machine will impose unacceptable performance penalties.

To solve this problem, we propose a *shared virtual memory* approach. Note that pointer aliasing becomes an issue because of the shared nature of the virtual address space from the perspective of intra-process migration. In our solution, we statically allocate regions of the virtual address space to participating processors. The first processor allocates dynamic memory in the region [1,X] MB, the second processor allocates memory in the region [X, 2X] MB and so on, where the memory addresses are in virtual memory space. In this model, when a string migrates from a source processor to a destination processor, it is guaranteed that memory addresses in its VM space are available on the target processor. This solution effectively bypasses the pointer aliasing problem[4].

The sharing of single VM address space across multiple processors imposes size restrictions on the VM addresses that can be allocated to any single processor, which in turn impacts the dynamic size of an application. However, this is not as restrictive as it appears at first sight. On a 64 bit processor, we can allocate 1 TB of VM space to each processor and still support parallel applications that can run on 16 million CPUs - well beyond the scope of current applications and super computers (the calculation divides the 64 bit space into 40 bit VM addresses and 24 bit CPU identifiers). As the dynamic memory demands grow, we can allocate more bits to the address space, reduce the maximum number of CPUs that can participate in the computation and still stay ahead of Moore's Law.

Performing a similar calculation for 32 bit processors shows that the above scheme imposes significant restrictions. For instance, if we allocate 1 GB of VM space per processor, we can only support applications that can run on 4 processors, which is definitely not acceptable. To get around this issue, we note that the main 32 bit processor family is based on the Intel x86 instruction set. Starting with the Pentium Pro family, Intel added 4 additional bits to the addressing, resulting in a 36 bit VM address space. This allows us to allocate the full 32 bit address space to each processor and support parallel applications up to 16 CPUs.

---

[4] We have also looked at dynamic VM partitioning schemes. Dynamic partitioning increases the cost of dynamic memory allocation calls. Hence, we propose a static partitioning scheme.

Even with the additional 4 bits of VM addressing space, limiting parallel applications to 16 CPUs is overly restrictive. To ameliorate this condition, we compartmentalize CPUs into *VM regions* of 16 CPUs each. Strings can freely migrate within a region and with some restrictions, even across regions. This solution offers an attractive trade-off between scalability and run-time load-balancing for distributed memory architectures.

# 4 Adaptivity Schemas

In working with concerted groups of scientists and engineers engaged in Grid computing, we have encountered a number of recurring "schemas" capturing how compositional scientific codes should be configured for adaptive execution. This section outlines these schemas and identifies application contexts where they are relevant.

Before we begin, it is pertinent to mention that two common modes of high-level Grid problem solving – viz. **parameter sweeps** [Casanova et al., 2000], **algorithmic bombardment** [Barret et al., 1996] – are easily supported using the Weaves framework. Parameter sweeps embody rich opportunities for state sharing and overloading of function invocations, and Weaves enables such sweeps to be conducted within an economy of processes. Offline recommender systems rely on the ability to conduct multi-dimensional parameter sweeps effectively and economically. Algorithmic bombardment is a speculative strategy by which multiple algorithms or solution approaches are assigned to a given problem (simultaneously), some of which may not run to completion and/or may be terminated when they are deemed redundant. Simplistically, algorithmic bombardment can be implemented efficiently through spatial adaptivity. From a simulation perspective, however, the end-goals of such bombardment can be achieved more elegantly through the notions of optimistic temporal and spatial adaptivity. Such a system will not be required to recover from any failures or revisit an earlier stage in the computation.

The list of adaptivity schemas below (see Table 1) is merely meant to be indicative of the power of our runtime systems framework and the coverage is not intended to be exhaustive.

Table 1: Adaptivity schemas currently supported in our research.

| Adaptivity Schema | Example Application Context |
| --- | --- |
| Staged Composition | Compositional PDE Solver Selection |
| Adaptation of Problem Decompositions | Numerical Quadrature, Adaptive Sorting |
| Coordinated Problem Solving | Interface Relaxation Algorithms |
| Algorithm Switching | ODEs, Number Factoring |
| Control Systems | Deriving Controllers for Algorithm Speedups |
| Active Mining of Recommendation Spaces | Qualitative Assessment for Matrix Computations |
| Graphs of Models | Multi-paradigm Performance Profiling |

## *4.1 Staged Composition*

Staged composition addresses the *sequential* selection and execution of code modules in scientific computations. It is important in problem domains that are characterized by *partial observability*. In this schema, code fragments from a library are composed at runtime to satisfy various general and domain-specific constraints on their structure. For instance, in the PDEs domain, the code fragments would correspond to choices of discretizer, pre-conditioner, and linear system solver. Since information about application performance characteristics is often acquired during the actual computation, rather than before, staged composition is a necessary feature in many application domains.

At each stage, the runtime recommender uses any of the features assessed and mined performance data to make a selection for a code module. In addition, the recommender can exploit a variety of considerations for staged composition: (i) domain-specific restrictions, (ii) interaction heuristics, and (iii) behavioral and performance characteristics. Domain-specific restrictions refer to both syntactic and semantic constraints on compositional modeling. An example of a syntactic restriction is that a compositional PDE solver must activate a discretizer, indexer, and linear system solver, in that order. A different permutation of these parts does not make syntactic sense. An example of a semantic restriction is that a Dyakunov algorithm requires that its input be in self-adjoint form. Interaction heuristics refer to considerations that bridge the various stages of the compositional process. Behavioral and performance issues are used to denote considerations such as "the main cost to solving a PDE is usually that of solving the linear system associated with it", and "Sharp ridges and other difficulties such as re-entrant corners cause difficulties in the estimation of convergence."

A runtime recommender can use such considerations to prune the search space of code modules and scale its functionality to large domains. In this *model-based approach,* the sequence of stages in a composition is captured using a Markov decision process and the utilities of states are directly estimated. Then, given an initial state, the runtime recommender would evaluate the various choices (of algorithm components) and choose the one that leads to the state with the highest utility.

## 4.2 Adaptation of Problem Decompositions

Many scientific computations are characterized by a recursive divide-and-conquer strategy, with algorithm selection happening at each level of the recursive invocation. Classical examples are adaptive numerical quadrature and adaptive sorting on parallel architectures. With the Weaves framework, the runtime recommender has the capability to backtrack both breadthwise and depthwise in the recursive function invocation history. This means that any form of branch-and-bound algorithm can be easily implemented. Notice that the breadthwise capability arises from the *parallel* compositional nature of the Weaves framework.

To curtail the potential explosive growth in space complexity, the runtime recommender must cleverly choose an intermediate representation that is indicative of the problem characteristics and, at the same time, can be cheaply evaluated when necessary. This is because at each backtrack point, the recommender has to make a judgement of code module and execution path. The choice of the intermediate representation is a domain-specific issue but we can give an indication of what it might look like. In the case of recommending numerical quadrature algorithms, it is of critical interest to assess features of the integrand such as the presence of a singularity, whether it is an end-point singularity, whether the integrand is smooth in the interval, and whether it exhibits an oscillatory behavior of non-specific type. These features are sometimes impossible to determine (e.g., when the integrand is provided only as a software routine). One solution approach is to first model the dynamic selection of quadrature nodes by a general purpose adaptive code such as QAGS [Piessens et al., 1983] and then use the layout of *these* nodes as the actual representation of the function. This requires that we employ optimistic temporal adaptivity in order to be able to successfully backtrack and later follow a suitable integration algorithm.

## 4.3 Coordinated Problem Solving

The collaborating PDE solvers application described earlier falls in this category. Here, adaptivity is the responsibility of one/some of the weaved code modules themselves (in this case, the mediators), and which coordinates the functioning of other code modules. Note that the structure of the composition – shared elements and multiple flows of control (see Fig. 1) - is naturally prone to single-cycle deadlock. While an implementation may be carefully instrumented to avoid deadlocks, the Weaves framework enables us to use the natural, underlying, problem representation and rely on runtime systems support for deadlock detection and recovery. The discussion section contains details of this mode of operation.

## 4.4 Algorithm Switching

Algorithm switching refers to the case where the problem being solved remains the same but the currently executing algorithm has to be replaced with another, dynamically. This facility is critical in solving ODEs with both stiff and non-stiff components, solving certain categories of linear systems, and integer factoring. For instance, the ODEs underlying many biological cell cycle models alternate between being stiff and non-stiff several times over the region of integration. In addition, properties such as stiffness are really a facet of both the ODE and the algorithm used to solve it. Algorithm switching is relevant here because our understanding of the problem improves as the computation proceeds. LSODE [Petzold, 1983] is an example of a real scientific code that embodies an algorithm switching mechanism, but as mentioned earlier the switching procedure is hardwired. It is sometimes "overcautious" to prevent thrashing between the two categories of algorithms. This is because, since stepsize selection is dependent on error estimates, situations involving misleading estimates can cause either a premature termination of methods or a switch to an unstable method. A runtime recommender can more carefully assess the suitability of algorithm switching by taking into account problem characteristics and runtime information, not otherwise available to the basic ODE algorithm.

In other applications, algorithm switching is important because the initial choice of algorithm fails. Here, it is imperative that we are able to use results and byproducts from the first algorithm to "seed" subsequent algorithm recommendations. For instance, in crypto-challenges such as integer factoring [Silverman and Wagstaff, 1993], we might switch to the quadratic sieve algorithm when the elliptic curve method fails.

## 4.5 Control Systems

An algorithm control system can be modeled with various configurations of the runtime recommender in the problem solving loop. More fundamentally, many classical formulations of control systems can be realized in scientific codes. For instance, a simple form of derivative-based control was used by Hovland and Heath [Hovland and Heath, 1997] to achieve an adaptive control policy for the SOR (Successive Over-Relaxation) algorithm. This is shown to be more powerful than using a fixed one with the optimal value of the over-relaxation parameter!

Similarly, adaptive control formulations are common in solving ODEs and automatic quadrature. In the former, the problem of stepsize selection can be thought of as designing a suitable controller (P, PI, PD, or PID formulations) around the basic numerical approximation. Automatic quadrature algorithms embody control systems because they must inherently assess the suitability of their approximations by deriving error estimates (often using approximations of successive orders).

A runtime recommender system extends such control system formulations into the realm of *actor-critic* models; the *actor* is the recommender that makes selections of solution components and the *critic* captures the improvement in how the recommender is itself assessed. Both the actor and the critic are implemented as learning algorithms. As the critic is learning to exercise better judgement, the actor benefits from the improved assessments, leading to a closed-loop control system.

## 4.6 Active Mining of Recommendation Spaces

In assessing many recommendation spaces, it is important to selectively sample and actively collect data, for the sole purpose of improving the confidence in the recommendation. For instance, in qualitative assessment of Jordan forms [Ramakrishnan and Bailey-Kellogg, 2002], data points are actively collected at specific perturbations in order to determine the most probable Jordan form of a matrix. This adaptivity schema iterates between a code execution (for collecting a data point), refining the recommendation (another code execution), and repeating these steps until a desired functional is minimized. This idea is a central ingredient of the US National Science Foundation's recent thrust for Dynamic Data-Driven Application Systems (DDDAS; [Darema, 2002]).

Another critical application arises in integrating data from measurements and data from performance evaluation of codes, to improve confidence in a model. A popular example is biological cell cycle models (e.g., for simulating the Golgi apparatus) where rate constants are collected from measurements (stored in files), and must be reconciled with simulation results (available from code executions).

## 4.7 Graphs of Models

In this final adaptivity schema, adaptivity is itself factored as operations on a graph and the task of runtime recommendation reduces to traversing this graph, to achieve user-specified criteria. For instance, in the performance modeling of the Sweep3D code ([Koch et al., 1992]; a benchmark for discrete-ordinates neutron transport), codes are available for analytical modeling, low-level simulation, and actual system execution [Adve et al., 2000]. Each node in the "graphs of models" corresponds to one model family, and the edges denote conditions and constraints to be satisfied (or achieved) when switching from one model to another. Consider two scenarios of Sweep3D modeling: one might (i) model the machine parameters accurately, taking into account processor components, memory components (buffers etc.) and transport components (interfaces to caches), or (ii) one might replace all machine parameters by picking one of the analytical models. Thus, moving from (i) to (ii) in the models graph might take place under the constraint that over 65% of the parts of the composed application need to be removed. Given end-to-end performance constraints, the runtime recommender then attempts to perform a means-end analysis on the induced graph, leading to a *satisficing model sequence*, that involves both models and the edges connecting them (notice that there may be more than one edge between two model choices). Preliminary results for this application are reported in [Houstis et al., 2002].

# 5 Early Results

## 5.1 Weaves: Implementation and Evaluation

The core of the Weaves compositional framework is the abstraction of a weave, which allows an application composer to define arbitrary namespaces over a composed application. To implement the weave abstraction, we need a data structure that can efficiently capture the state separation and state recombination needs of the compositional framework.

Before we discuss the specifics, note that the goals of our compositional framework place additional constraints on the implementation of the weave abstraction. First, our transparency requirement states that the solution should be transparent to the application. Since, the application may be written in any programming language, the transparency requirement precludes modification to the source code to implement the namespace abstraction. Second, from a scalability perspective, the implementation should be efficient. In particular, we need to minimize context switch time between the various namespaces defined in the composed application.

To meet the transparency requirement, the implementation of the namespace abstraction works by analyzing the Executable and Linking Format (**ELF**) object files produced by any compiler. ELF is a public domain file format used to represent both object code as well as the final executable on most UNIX systems. Our current prototype is implemented on the Linux operating system running on Intel x86 architectures. Since the implementation only depends on the ELF file format, it can be easily ported to other operating systems/architectures. Furthermore, we anecdotally note that the features of the ELF file format used by our implementation are common to object file formats. Hence, it should be possible to extend the prototype to support other object file formats as well.

The ELF file format uses the Global Offset Table (**GOT**) data structure to access global state in an application. The GOT data structure maintains an array of *pointers* (instead of data values), with each pointer referring to a global data variable. To access data, applications first index into the GOT data structure to get a pointer to the data and then use the pointer (and possibly an offset) to retrieve the data value. The number of entries in the GOT structure is proportional to the number of variables and is independent of the size of each variable. For instance, an array variable has a single entry in the GOT structure. The observation here is that the GOT defines the namespace of the application. Typically, a program contains a single GOT structure reflecting the single namespace within an application. However, by appropriately defining multiple GOT structures, it should be possible to create multiple namespaces within a single ELF executable.

The problem with the basic GOT structure is that compilers hardcode the base address of the GOT structure and the index into the GOT at compile time. To implement multiple namespaces, we need to create multiple GOT structures and, at runtime, copy them over to the fixed base address generated by the compiler. This operation is expensive since its cost is proportional to the number of global variables, which can potentially be large.

Instead, we note that compilers produce relocatable code (for instance, the command line option –fPIC on the gcc family of compilers) to support dynamic libraries. In relocatable codes, the base of the GOT structure is pointed to by a base register. All indexed accesses into the GOT are made with reference to the current value of the base register. The use of relocatable object code and indexed access to the GOT forms the basis of our implementation.

To implement the weave abstraction, we create a new GOT structure for each distinct weave in the composed application. To implement state separation between beads belonging to different weaves, we first create copies of the data and point the GOT entries in the weaves to the distinct copies of the data. To enable state recombination between weaves sharing a bead, we set the pointer in the GOT entries in the different weaves to point to the same data value. The double indexed nature of the GOT structure enables state separation/recombination at the resolution of a single data variable, which can be used to implement arbitrary data sharing at both the tuple space and module levels.

To implement the string abstraction, note that a string is really a thread operating under a user specified namespace. Since we have a mechanism to create the namespace, context switching between strings involves context switching the thread state and switching the namespace. What we need here is an efficient mechanism for switching namespaces.

To switch namespaces, we note that the GOT structure is accessed through a base register (%ebx in our current implementation). Hence, context switching between namespaces merely involves changing the base register to point to a different GOT structure, a single instruction move operation, which results in a weave context switch time that is identical to thread context switch time. Our current implementation of the Weaves framework works over both POSIX Threads (pthreads) as well as the GNU Portable Threads (Pth) thread libraries.

To implement spatial adaptivity, we need mechanisms to (i) dynamically insert/remove functions and (ii) implement dynamic function overloading. In addition, for flexibility reasons, spatial adaptivity should operate at the granularity of a single weave - beads in distinct weaves may have different implementations of a similarly named function.

Dynamic code insertion/removal can be implemented through a runtime interface to dynamic libraries, which "weaves" them on-the-fly. To see how (ii) can be implemented, notice that it is similar to a runtime version of latebinding virtual functions in OOP. Such dynamic function overloading can be used to transparently change functional implementations. The Weaves framework achieves this effect through an interesting use of the GOT namespace. Traditionally, dynamically linked functions are referenced through a single procedure linkage table (**PLT**). Our goal of spatial adaptivity requires multiple PLT structures, with each PLT referring to functional bindings in a single weave.

Instead of modifying the memory image of an executable to create multiple PLT structures, we chose to direct the compilation process through an intermediate assembly code generation stage. Here, we patch the assembly code to indirect functional references through the GOT, instead of the PLT. The initial PLT is folded into the GOT structure. This process extends the Weaves namespace abstraction to include both code and state, which is almost identical to the OOP view of an object. The namespace view allows us to change the function pointer in each distinct GOT enabling namespace specific function bindings. This mechanism achieves both dynamic function binding as well as primitive function overloading capabilities through weave specific function signatures.

We ran a series of experiments to compare the context switch time under the threads, processes and weaves programming models. In this experiment, we created a baseline application that implements a calibrated delay loop (busy wait). We then implemented threads-, processes-, and weaves- versions of the application. In each of these versions, there are $n$ independent flows of control over the same code, where each flow of control executes a calibrated delay loop, which does $1/n^{th}$ the work of the baseline application. We then measure the total time taken to execute the application under each of these models. Since each of the control flows does $1/n^{th}$ of the work and there are $n$ flows, the total time taken should the same as the baseline calibrated delay loop case, except for an additional context switching cost.

Figure 9 shows the results of the experiment on a single processor AMD Athlon™ workstation running the Linux operating system. The results show the run time for five cases: (a) baseline calibrated delay loop, (b) pthreads threads library, (c) Pth threads library, (d) processes, (e) Weaves over pthreads, and (f) Weaves over Pth. The results clearly show that the weaved implementations are significantly faster than processes, even in this simple case, where the copy-on-write semantics of the *fork()* call are very effective. Furthermore, the run time of weaved implementation of pthreads is very close to the base run time of pthreads alone. The marginal variation in runtime is due to the slightly higher weave creation cost, which is included in the run time. Also, the pthreads implementation is relatively efficient, since the Linux kernel includes operating system support for it.

However, in the case of Pth, the run time of the weaved implemented is higher than the base Pth case. This increase in runtime is because unlike pthreads, Pth is a user-level library and hence suffers from timer inaccuracies inherent in user-level library implementation.

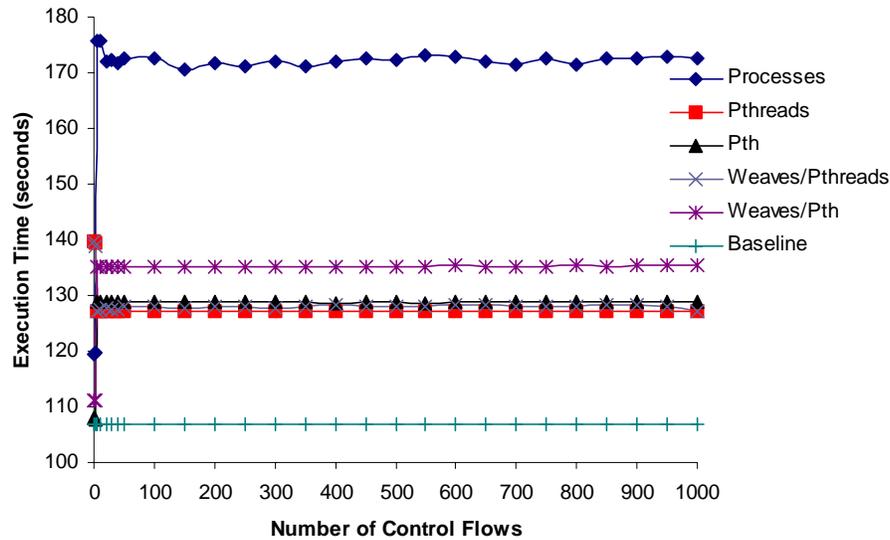

**Figure 9:** Comparison of inter-flow context switch time in the threads, processes, and weaves programming models. The baseline single process application implements a calibrated delay loop of 107 seconds.

## 5.2 Experiments in Adaptive Runtime Composition

A number of scientific applications have been or are currently being created using the runtime systems support framework described in this paper. These include:
- A weaved version of the Sweep3D code suitable for performance characterization over the Grid
- Compositional PDE solvers for multi-domain, multi-physics problems
- A runtime recommender system for adaptive numerical quadrature
- Iterative assessment of spectral portraits of matrices by active mining
- Adaptive ODE algorithm switching for simulating biological cell cycle models
- Dynamic selection of linear system solvers for molecular electronics simulations

Due to space considerations, we describe only the first application below, which embodies the "graphs of models" adaptivity schema described earlier. The role of a recommender system for Sweep3D characterization is well motivated in [Adve et al., 2000]; preliminary results for a recommender built on this idea are presented in [Houstis et al., 2002]. To avoid duplication, we focus here on how the Sweep3D application has been weaved and an assessment of its performance characteristics. The reader should keep in mind the larger context in which such a performance model is then used to drive the characterization of large-scale scientific applications.

### 5.2.1 Sweep3D

The main characteristic of Sweep3D is that it uses no global variables. Since the application only relies on local state, multiple instantiations of local state should be enough to create a VM abstraction. This characteristic makes Sweep3D inherently thread-safe, which enables its modeling by either the threads or process models. However, since the application is written in Fortran 77, with dynamic array extensions, modeling with the threads and processes models present interesting implementation problems. *While trying to model the application using POSIX threads, we found that there was substantial global state in the .data section of the ELF executable, a Fortran compiler issue, which essentially made the code-base "thread unsafe".* Weaving the Sweep3D code-base created independent namespaces, resulting in a thread-safe version.

To support the message passing primitives used by Sweep3D, we created a simple threaded MPI emulator, which implements only the **nine** MPI primitives used by Sweep3D. To ensure correctness, the MPI emulator implementation follows the guidelines set forth in the MPI specification. Our MPI emulator is intended as a test prototype and is neither as comprehensive nor as capable as a complete MPI implementation.

In the weaved implementation, we create *n* distinct virtual machines, each of which executes an independent instantiation of the Sweep3D application. To do this, we create *n* distinct Sweep3D beads and *n* weaves, where each weave has a distinct Sweep3D bead and a shared emulator bead. Each weave also has a single string associated with it. The *n* distinct virtual machines run on a single processor workstation.

We compared the performance of our single processor weaved implementation of Sweep3D against measured values from real runs for up to 150 processors. Measurements for the real runs were made on our 200 processor cluster (1GHz AMD Athlon ™ processors over Myrinet™) *Anantham*. Since the Sweep3D application performs its own timing measurements, we compared the timing numbers (CPU Time) of the weaved version of Sweep3D with the measurements from actual runs. The two input files (50x50x50 and 150x150x150 decompositions) provided in the Sweep3D distribution were used to drive the Sweep3D application.

For upto 150 processors, the timing results from the weaved implementation and the actual runs were consistent to within 0.2%. Furthermore, we tested the weaved version of Sweep3D with over 1000 weaves on a single processor. The variation in the timing results between multiple runs was within 0.2%. This clearly shows that even at high levels of scalability (over 1000 weaves/processor) context switch time does not impact the efficacy of our runtime compositional framework.

# 6   Discussion

This paper has described a novel runtime compositional system for supporting adaptive scientific computations on the Grid. Weaves serves as a true generalization of the threads and processes models of programming and provides immediate benefits in object-based composition, checkpointing, migrating, and dynamic reconfiguration of scientific applications. Runtime recommender systems encapsulate knowledge about which solution components perform well (and for which situations) and provide intelligent decision support for configuring and managing large-scale computations. Together, they constitute a powerful mode of developing and deploying adaptive grid applications.

The work presented here has interesting parallels to research in many different areas – we survey a collection of references topically. At a basic level, Weaves's capabilities as a programming model can be compared to that of distributed OO [Gannon and Grimshaw, 1998], parallel programming primitives [Foster, 1996; Skillicorn and Talia, 1998], agent-based composition [Drashansky et al., 1999], and service-based systems integration [Foster et al., 2002; Rana and Walker, 2001]. The design of the Weaves system bears a strong resemblance to the OO framework of Mentat propounded in [Grimshaw et al., 1996]. However, unlike Mentat, which requires creating code objects in an OO language, Weaves can create an object based framework from code written in any language, allowing the reuse of the vast repository of legacy codes.

In any parallel compositional framework of the type presented here, deadlock detection and recovery pose serious concerns. This issue is particularly problematic for us due to the dynamic nature of our framework, which prevents *a priori* analysis of deadlock scenarios. In ongoing research, we are working on implementing automatic mechanisms for transparent detection and elimination of single cycle deadlocks. The basic mechanism works by implementing functional continuations of mutual exclusion calls. Before acquiring a mutual exclusion lock, the continuations automatically (i) associatively track the strings and the bead invoking the mutual exclusion lock and (ii) checkpoint the beads in the string invoking the mutual exclusion call. Single cycle deadlock can then be detected through cycles in the history of acquired mutual exclusion locks. The actual detection mechanism can be implemented through a checkpoint extension to the monitor interface, which itself can be guaranteed to be deadlock free. In this setup deadlock recovery reduces to (i) choosing a candidate *victim string* and (ii) rolling back the victim string to its checkpoint state just prior to acquiring the lock.

Significant research has also been conducted to realize adaptivity in distributed scientific computations are well studied, e.g., in the contexts of performance modeling [Vraalsen et al., 2001; Adve et al., 2002], application tuning [Chang and Karamcheti, 2001], and meta-modeling and control [Ribler et al. 2001; Kennedy et al., 2002]. Many of these applications are focused on selecting system configurations, identifying optimal application parameters, and exploiting opportunities for application scheduling over the Grid. The notion of runtime recommendation presented here applies more broadly to selecting algorithms and code modules, and the knowledge-based framework allows application-specific context about the suitability of algorithms to be exploited. The algorithmic framework used for runtime recommendation (namely, reinforcement learning) is very powerful, and is part of a larger family of strategies for adaptive control of algorithm executions.

As Grid infrastructure improves and newer applications are explored, we believe the importance of Grid programming primitives will be better appreciated. It will be especially crucial that the programming primitives allow rich forms of adaptivity to be specified and captured without the need for low-level system configuration. There are many recent steps taken in this direction (e.g., the compiler directed frameworks described in [Adve et al., 2001; Adve and Sakellariou, 2000]). The central idea here is to encode adaptivity as operations on a suitably defined *task graph*, which serves as an intermediate representation of the dynamic behavior of a grid application. In addition to operationalizing adaptivity, such a representation allows systematic performance characterization of scientific applications using multiple methodologies [Browne et al., 2000]. In our work, the intermediate representation is the purview of the runtime recommender but dynamic operations of spatial and temporal adaptivity are handled by the checkpointing and composition framework supplied by Weaves. We are currently in the process of defining a language for declaring search primitives (akin to a branch-and-bound operation for optimistic simulation) that can be used as building blocks of adaptivity. The advantage with this formulation is that adaptivity is taking place at the level of code modules and hence can be made as coarse or fine grained as necessary. It also allows for ease of specification by the grid application developer.

The eventual success of grid computing will lie in "what it lets you get away with." By factoring support for adaptivity in a runtime recommender system and operationalizing parallel composition, checkpointing, and migration using the Weaves framework, the ideas presented here allow us to transparently realize the promise of adaptive Grid applications.

## *Acknowledgements*

This work is supported in part in US National Science Foundation grants EIA-9974956, EIA-9984317 (CAREER), EIA-0103660, and EIA-0133840 (CAREER).